\documentclass[a4paper,10pt,twocolumn]{article}
\usepackage[utf8]{inputenc}
\usepackage{multicol,graphicx}
\usepackage{mathptmx}
\usepackage{bm} 
\usepackage[T1]{fontenc}
\usepackage{textcomp}
\usepackage{url}
\usepackage[hidelinks]{hyperref}
\usepackage[T1]{fontenc}
\usepackage[english]{babel}
\usepackage{booktabs}

\usepackage[backend=bibtex,style=ieee,doi=false,isbn=false,url=true,maxnames=6,citestyle=numeric-comp,giveninits=true]{biblatex}

\fontfamily{ptm}\selectfont

\bibliography{references.bib}

\usepackage[font=small]{caption}
\usepackage{amsmath, amssymb}

\newcommand{\mysection}[1]{\vspace{0.4cm} \uppercase{#1} \vspace{0.4cm}}
\newcommand{\mysubsection}[1]{\hspace{10pt}\textit{#1:}}

\begin{document}
	
\setlength{\textfloatsep}{10pt plus 1.0pt minus 2.0pt}	
\setlength{\columnsep}{1cm}


\twocolumn[%
\begin{@twocolumnfalse}
\begin{center}
	{\fontsize{14}{18}\selectfont
        \textbf{\uppercase{
        Exploring new territory: \\calibration-free decoding for c-VEP BCI}}\\}
    \begin{large}
        \vspace{0.6cm}
        J. Thielen\textsuperscript{1}, J. Sosulski\textsuperscript{2}, M. Tangermann\textsuperscript{1}\\
        \vspace{0.6cm}
        \textsuperscript{1}Donders Institute, Radboud University, Nijmegen, the Netherlands\\
        \textsuperscript{2}Department of Computer Science, University of Freiburg, Freiburg, Germany\\
        \vspace{0.5cm}
        E-mail: \href{mailto:jordy.thielen@donders.ru.nl}{jordy.thielen@donders.ru.nl}
        \vspace{0.4cm}
    \end{large}
\end{center}	
\end{@twocolumnfalse}%
]%


ABSTRACT: 
This study explores two zero-training methods aimed at enhancing the usability of brain-computer interfaces (BCIs) by eliminating the need for a calibration session. We introduce a novel method rooted in the event-related potential (ERP) domain, unsupervised mean maximization (UMM), to the fast code-modulated visual evoked potential (\mbox{c-VEP}) stimulus protocol. We compare UMM to the state-of-the-art \mbox{c-VEP} zero-training method that uses canonical correlation analysis (CCA). The comparison includes instantaneous classification and classification with cumulative learning from previously classified trials for both CCA and UMM. Our study shows the effectiveness of both methods in navigating the complexities of a \mbox{c-VEP} dataset, highlighting their differences and distinct strengths. This research not only provides insights into the practical implementation of calibration-free BCI methods but also paves the way for further exploration and refinement. Ultimately, the fusion of CCA and UMM holds promise for enhancing the accessibility and usability of BCI systems across various application domains and a multitude of stimulus protocols.


\mysection{introduction}

A brain-computer interface (BCI) records the user's brain activity and converts these into computer commands, offering an alternative output channel that does not rely on muscular activity. Electroencephalography (EEG) is commonly used to record brain activity due to its affordability, practicality, and non-invasiveness. The primary application of BCIs lies in restoring lost control, particularly in communication. One notable example is the visual BCI speller, where users can select symbols by focusing their gaze on them when displayed on a screen. This technology proves invaluable for individuals facing challenges such as amyotrophic lateral sclerosis, where progressive loss of voluntary motor control makes speaking and typing difficult~\cite{verbaarschot2021}. 

Prior to BCI usage, a machine learning model capable of classifying unseen brain signals needs to be calibrated on labelled EEG data from the same user, as individuals display different patterns of brain activity. Additionally, the same user might show different patterns over multiple days of use (session-to-session variability) and even within-session non-stationarity. To mitigate any negative effects of these confounders, the user is guided through an initial stage to record brain activity while being instructed which symbol to attend to. 

While a trained classification model is necessary for using the intended BCI application, the calibration recording delays a deployment and may be prohibitive specifically for users with a limited attention span. In general, the necessity of calibration may impede the acceptance and widespread adoption of BCIs by patients and healthy users. Encouragingly, recent advancements in BCI technology have surfaced which offer potential solutions to alleviate this challenge.

The first advancement involves selecting an informative brain signal feature to minimize the duration of the calibration phase. BCIs can be driven by various brain signals, often evoked by advanced stimulus protocols. Many popular stimulus protocols induce one or more event-related potentials (ERPs). Among these, the visual evoked potential (VEP), triggered by a flash, stands out. VEP-based BCIs are widely embraced due to their effectiveness across diverse user populations~\cite{volosyak2020}. The VEP can be effectively used in three different ways~\cite{bin2009}. 

Firstly, in a BCI based on time-modulated VEP (t-VEP), stimuli are sequentially presented to reduce temporal overlap, resulting in a relatively slow paradigm. Secondly, in a frequency-modulated (f-VEP)-based BCI, each stimulus simultaneously and rapidly flashes at a unique frequency and phase~\cite{nakanishi2017}. Despite its speed, f-VEP faces limitations due to the restricted range of narrow-band options and potential artefacts that may obscure signals. Thirdly, in a code-modulated (\mbox{c-VEP})-based BCI, each stimulus rapidly flashes with a pseudo-random noise-code~\cite{martinez2021}. In this protocol, stimulus sequences are optimized to be dissimilar, ensuring that their corresponding brain activity is dissimilar as well. Remarkably, \mbox{c-VEP} BCI has recently demonstrated unprecedented performance~\cite{shi2024}.

The second advancement involves the choice of the decoding approach. In the case of \mbox{c-VEP}, a method was developed, termed `reconvolution', which relies on a forward model embedded in a canonical correlation analysis (CCA)~\cite{thielen2015}. This model characterizes the response to a sequence of flashes as the linear summation of responses to individual flashes. This reconvolution approach substantially reduced the number of trainable parameters while it simultaneously increased the number of samples that were available as training data points. This not only decreased the required training data but also empowered the model to predict responses to unseen stimulus sequences. Recently, this reconvolution CCA method was shown to achieve remarkable performances on \mbox{c-VEP} data event without the need for a calibration session, by finding the stimulus sequence that best fits the data in a trial~\cite{thielen2021}.

In a recent development, a novel classification approach was introduced for ERP-based BCI. The approach, termed `unsupervised mean-difference maximization' (UMM) does not require labeled calibration data~\cite{sosulski2023}. UMM does not act on every single epoch, i.e., the evoked response of a single stimulus, but on a set of epochs belonging to one control command, e.g., one trial that leads to the selection of a symbol in a spelling application. Therefore, UMM has similarities to the aforementioned CCA method. While the objective of CCA is to maximize the explained variance, UMM identifies an attended target symbol by maximizing the distance between target and non-target ERPs belonging to one trial. Diverging from the CCA method, UMM incorporates several regularization approaches, including the use of block-Toeplitz covariance matrices~\cite{sosulski2022} for determining domain-specific distances, and it can take advantage of a built-in confidence metric. Collectively, UMM has demonstrated impressive performance across various ERP datasets, without the need for a training session~\cite{sosulski2023}.

In this study, we aim to combine the efficiency of the \mbox{c-VEP} stimulus protocol and the carefully regularized UMM approach for zero-training. This will be the first instance that UMM is applied to \mbox{c-VEP} data, arguably a much faster stimulus protocol than the conventional ERP stimulus protocol. In the analysis, we draw a comparison with the CCA zero-training pipeline that was already evaluated on \mbox{c-VEP} data. This study not only sheds light on the efficacy of CCA and UMM, but also deepens our understanding of their underlying distinct mechanisms, unraveling insights into constructing effective BCIs for communication and control. By eliminating the need for a calibration session, this research paves the way for plug-and-play BCIs, marking a significant stride towards user-friendly and accessible BCI technology.


\mysection{materials and methods}

\mysubsection{Dataset}
We assessed the efficacy of the CCA and UMM zero-training approaches using an open-access \mbox{c-VEP} dataset~\cite{thielen2023dataset}. Comprehensive details about this dataset can be found in the original study~\cite{thielen2021}. For the current study, we only used the part of this dataset labeled as `offline experiment', in which 30 participants engaged in a copy-spelling task. EEG data were recorded from 8 electrodes placed following the 10-10 system (Fz, T7, O1, POz, Oz, Iz, O2, T8) amplified using a Biosemi Active2 amplifier and sampled at a frequency of 512\,Hz. 

Throughout the experiment, participants interacted with a $4 \times 5$ matrix speller displayed on a 12.9\,in iPad Pro with a 60\,Hz refresh rate and a $1920 \times 1080$\,px resolution. The $N=20$ cells within this matrix each measured 3.1\,cm $\times$ 2.8\,cm, with a 0.4\,cm separation both horizontally and vertically between cells. The cells were presented against a mean-luminance gray background. Each cell $i \in \{1, \dots, N\}$ was luminance modulated using a unique binary stimulus sequence at full contrast, with a 1 encoding a white cell and a 0 a black cell. The stimulus sequences were carefully chosen from an optimized subset of Gold codes~\cite{gold1967}. These sequences were modulated such that they contained flashes of only two durations: a short flash of 16.67\,ms and a long flash of 33.33\,ms. The sequences had a length of $126 = 2 * (2^6 - 1)$ bits and at 60\,Hz cycling through a stimulation code once took 2.1\,s. 

Participants completed 5 identical runs, with each run comprising 20 trials, one for each of the 20 cells presented in a random order. Each trial started with a 1-second cue highlighting the target cell in green. Subsequently, all cells started flashing with their respective stimulus sequences for a duration of 31.5\,s (equivalent to 15 code cycles), during which participants maintained fixation on the target cell. Post-trial, no feedback was provided and the subsequent trial commenced without delay. To sum up, each participant contributed 100 trials of 31.5\,s, including 5 repetitions for each of the 20 stimuli.

The EEG data underwent preprocessing using Python version\,3.10.9 and MNE version\,1.6.0. Initially, a notch filter at 50\,Hz was applied to eliminate line noise. This was followed by a band-pass filter with a lower cut-off at 6\,Hz and an upper cut-off at 50\,Hz, which was optimized in an initial analysis. Subsequently, the data were segmented into single-trials, spanning from 500\,ms before stimulus onset to 31.5\,s after stimulus onset. The dataset was then downsampled to 180\,Hz, which is a multiple of the monitor refresh rate at 60\,Hz. Finally, the initial 500\,ms of data per trial, which may have caught artefacts resulting from the initial slicing and subsequent filtering processes, were removed. 

\mysubsection{Canonical correlation analysis (CCA)}
Using CCA, let's assume that the current trial $\mathbf{X} \in \mathbb{R}^{C \times T}$ contains $T$-many temporal features extracted from each of the $C$-many channels. Here, $C=8$. To decode the attended target symbol $\hat{y}$ of a new trial via CCA, each of the $i \in \{1, \dots, N\}$ possible hypotheses about which cell, i.e., which stimulus sequence, may have represented the target, are considered. Here, $N=20$. 

As CCA operates at the trial level, its $i$th stimulus sequence is described by the event time-series $\mathbf{E}_i \in \mathbb{R}^{E \times T}$ for $E$-many events and $T$-many temporal features. Here, we modeled $E=3$ events including the two flash durations and an onset event for the sudden start of the stimulation.

Subsequently, the event time-series are transformed into a structure matrix $\mathbf{M}_i \in \mathbb{R}^{M \times T}$ with $M$-many event time-points and $T$-many temporal features. Let's assume equally long responses to each of the $E$ events, then $M=E * L$. Here, $L=54$, which corresponds to 300\,ms at 180\,Hz. This structure matrix is a Toeplitz matrix describing the onset, duration and overlap of the responses to each of the events in the $i$th stimulation sequence.

We fit a CCA model for each of the $N$ candidate stimulus sequences $i \in \{1, \dots, N\}$ by learning sequence-specific spatial filters $\mathbf{w}_i \in \mathbb{R}^{C}$ and temporal filters $\mathbf{r}_i \in \mathbb{R}^M$:
\begin{equation}\label{eq:cca_fit}
    \arg\max_{\mathbf{w}_i, \mathbf{r}_i} \frac{\mathbf{w}^\top_i\mathbf{X} \mathbf{M}^\top_i\mathbf{r}_i}{\mathbf{w}^\top_i\mathbf{X}\mathbf{X}^\top\mathbf{w}_i\mathbf{r}^\top_i\mathbf{M}_i\mathbf{M}^\top_i\mathbf{r}_i}
\end{equation}

Instantaneous classification of the current trial, i.e., determining the one attended target symbol $\hat{y}$ from the $N=20$ symbols, is then performed by maximizing the correlation, which is equivalent to the square root of the explained variance:
\begin{equation}\label{eq:cca_predict}
    \hat{y} = \arg\max_{i} \frac{\mathbf{w}^\top_i\mathbf{X} \mathbf{M}^\top_i\mathbf{r}_i}{\mathbf{w}^\top_i\mathbf{X}\mathbf{X}^\top\mathbf{w}_i\mathbf{r}^\top_i\mathbf{M}_i\mathbf{M}^\top_i\mathbf{r}_i}    
\end{equation}

Alternatively to this instantaneous CCA, CCA can learn across trials. Specifically, previous trials can be included to improve the estimates for the current trial, as described in~\cite{thielen2021}. Equation~\ref{eq:cca_fit} can be formulated using the spatio-temporal cross-covariance ${\bm\Sigma}_{\mathbf{X}\mathbf{M}_i} \in \mathbb{R}^{C \times M}$, the spatial covariance ${\bm\Sigma}_\mathbf{X} \in \mathbb{R}^{C \times C}$ and the temporal covariance ${\bm\Sigma}_{\mathbf{M}_i}$:
\begin{equation}\label{eq:cca_fit_cov}
    \arg\max_{\mathbf{w}_i, \mathbf{r}_i} \frac{\mathbf{w}^\top_i{\bm\Sigma}_{\mathbf{X}\mathbf{M}_i}\mathbf{r}_i}{\mathbf{w}^\top_i{\bm\Sigma}_\mathbf{X}\mathbf{w}_i\mathbf{r}^\top_i{\bm\Sigma}_{\mathbf{M}_i}\mathbf{r}_i}
\end{equation}
The estimation of these covariance matrices can be improved by accumulating the data $\mathbf{X}$ and predicted structure matrix $\mathbf{M}_{\hat{y}}$ of the previous trial(s). This cumulative CCA is an optimistic one, as it assumes that previous trials were classified correctly (i.e., naive labeling).

In summary, in this work, we applied two versions of CCA. For both we used the empirical covariance matrix identical to the original work~\cite{thielen2021}. The first version, denoted CCA\_e1, was instantaneous and estimated the covariance from the current trial only. The second version, denoted CCA\_ec, was cumulative and used previous trials for covariance estimation to facilitate decoding of the current trial. 
Code for the CCA approach is available at \url{https://github.com/thijor/pyntbci}.

\mysubsection{Unsupervised mean-difference maximization (UMM)}
Using UMM, we first slice the current trial into the contained $K$-many epochs, which are synchronized to each bit in the stimulus sequences, i.e., the monitor refresh rate at 60\,Hz. 

Let's assume that an epoch $\mathbf{x} \in \mathbb{R}^D$ is described by a $D$-dimensional feature space, which contains $T$-many temporal features extracted from each of the $C$-many channels, i.e., $D=C*T$. Here, $C=8$ and $T=54$ for epochs of 300\,ms long at 180\,Hz. To decode the attended target symbol $\hat{y}$ via UMM for the current trial, each of the $i \in \{1, \dots, N\}$ possible hypotheses about which cell may have represented the target, are considered. Here, $N=20$. 

For every possible hypothesis $i \in \{1, \dots, N\}$ we then estimate the mean-difference vector $\Delta{\bm\mu}_i \in \mathbb{R}^{D}$, which is the difference between the flash ERP and non-flash ERP:
\begin{equation}\label{eq:umm_fit}
    \Delta{\bm\mu}_i = \frac{1}{|A^+_i|}\sum_{j \in A^+_i}\mathbf{x}_j - \frac{1}{|A^-_i|}\sum_{j \in A^-_i}\mathbf{x}_j    
\end{equation}
where $\mathbf{x}_j \in \mathbb{R}^{D}$ is the $D$-dimensional EEG feature vector of the $j$-th epoch, and $A^+_i$ and $A^-_i$ denote the sets of epochs for which a flash was either presented (bit is 1) or not (bit is 0) under the current hypothesis of $i$ being the target stimulus sequence.

Instead of determining the attended symbol of the current trial by maximizing the Euclidean distance between flash and non-flash ERPs across all $N$-many hypotheses, the metric is first normalized using the inverse of the global feature covariance matrix ${\bm\Sigma} \in \mathbb{R}^{D \times D}$ to better cope with non-spherical feature distributions in the feature space $\mathbb{R}^D$, known as the Mahalanobis distance:
\begin{equation}\label{eq:umm_predict}
    \hat{y} = \arg\max_{i} (\Delta{\bm\mu_i}){\bm\Sigma}^{-1}(\Delta{\bm\mu_i})
\end{equation}

Please note that the covariance matrix can be estimated based on the epochs of the current trial only, which makes UMM an instantaneous decoding approach that does not require calibration data. Due to the challenging ratio of the feature dimensionality and the number of epochs contained in a single trial, we used a block-Toeplitz regularization with tapering to obtain a more robust estimate of the covariance matrix~\cite{sosulski2022}. 

Alternatively to this instantaneous use of UMM, know\-ledge from previous trials about both, the estimated class means and the covariance matrix can be included for obtaining improved estimates for the current trial, as described in~\cite{sosulski2023}. Specifically, as the covariance matrix can be calculated without label information the covariance matrix can more robustly be estimated by also using epochs from previous trials~\cite{sosulski2023}. A similar approach can be used to more robustly estimate the flash and non-flash ERPs, by using information from previous trials. However, for this, label information is required. UMM simply uses its own predictions from previous trials as pseudo labels (sometimes also referred to as naive labeling). This approach is made more robust, by weighting the mean estimates from previous trials by UMM's confidence in these previous trials~\cite{sosulski2023}. Specifically, if UMM is very certain of its own prediction, these ERP means will have more weight in later trials and vice versa. 

\begin{figure*}[ht]
    \centering
    \includegraphics[width=\textwidth]{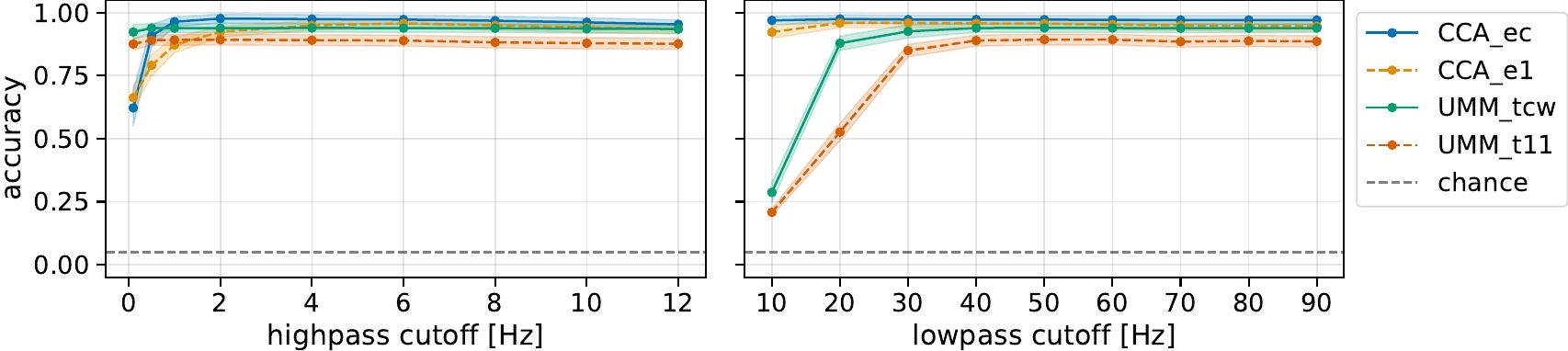}
    \caption{\label{fig:bandpass}\textbf{Bandpass hyper-parameters for CCA and UMM.} Depicted are the grand average classification accuracy for CCA and UMM across varying highpass (left) and lowpass (right) cutoff values. Here, a single-trial duration of 31.5\,s is used. When varying the highpass, the lowpass remained at 40\,Hz, and when varying the lowpass, the highpass remained at 6\,Hz. In order, the symbols behind a method refer to: the type of covariance matrix being empirical (e) or block-Toeplitz (t); covariance matrices computed instantaneously (1) or cumulative (c); and the mean vectors (of UMM) computed either instantaneously (1) or using a weighted cumulative average (w). The dashed gray line denotes the theoretical chance level (5\%).}
\end{figure*}

In summary, for this work, we used two versions of UMM. For both versions we used the block-Toeplitz regu\-la\-rized covariance matrix. The first version, denoted UMM\_t11, is instantaneous by estimating the covariance and means only from the current trial. The second version, denoted UMM\_tcw, is cumulative by using previous trials to facilitate decoding of the current trial. For an overview of all method's abbreviations used, please refer to the legend of Fig.~\ref{fig:bandpass}.
Code for the UMM approach is available at: \url{https://github.com/jsosulski/umm_demo}.

\mysubsection{Analysis}
This study assessed the effectiveness of CCA and UMM in classifying \mbox{c-VEP} data without calibration, including both an instantaneous approach, where trials are classified without any prior calibration (CCA\_e1, UMM\_t11), and a cumulative approach, where trials are classified while leveraging information from previously analyzed trials (CCA\_ec, UMM\_tcw).

As previously mentioned, we first varied the cutoff frequency of the highpass and lowpass filter in the bandpass spectral filter used for preprocessing, to explore the influence of these hyper-parameters on the performance of the analyzed methods. For the highpass we tested 0.1, 0.5, 1.0, 2.0, 4.0, 6.0, 8.0, 10.0 and 12.0\,Hz, for a total of 9 evaluations, all with a lowpass at 40\,Hz. For the lowpass we tested 10 to 90\,Hz in 10\,Hz increments, for a total of 9 evaluations, all with a highpass at 6\,Hz. These evaluations were always carried out using the full trial durations of 31.5\,s. Recognizing that each method may respond uniquely to these variations, our final comparative analysis focused on the bandpass cutoff frequencies that resulted in the highest classification accuracy.

To assess the methods, we generated decoding curves by varying trial durations from 1.05\,s (half a code cycle) to 10.5\,s (5 code cycles) in 1.05\,s increments, and from 10.5\,s to 31.5\,s in 2.1\,s increments, for a total of 20 decoding time steps. Across these time steps, the number of bits ranged from 63 to 630 in 63-bit increments, and 630 to 1.890 in 126-bit increments, directly corresponding to the available epochs for UMM at each decoding time step. Because modulated Gold codes have an equal number of ones and zeros, the number of flash and non-flash epochs were always equal or deviated at most by 1 when using half a code cycle, for any of the stimulus sequences. In this analysis, a bandpass filter of 6--50\,Hz was used, as it turned out optimal for all methods. 

In this study, all statistics were carried out using a one-sided paired Wilcoxon signed-rank test to test for a larger classification accuracy of one versus another method. The significance level was set to $\alpha=0.025$. Reported $p$-values were not corrected for multiple comparisons.


\mysection{results}
\begin{figure*}[h]
    \centering
    \includegraphics[width=\textwidth]{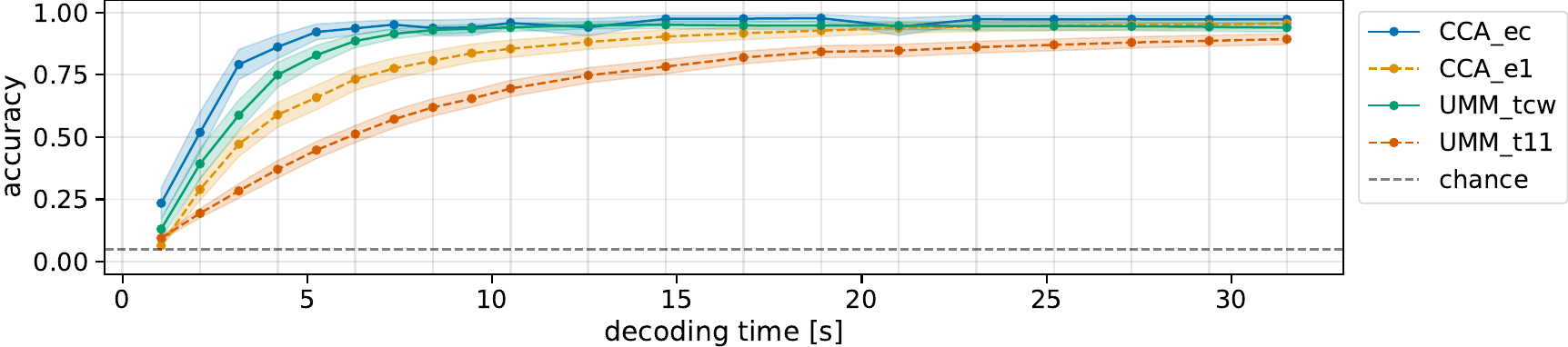}
    \caption{\label{fig:decodingcurve}\textbf{Decoding curve for CCA and UMM.} Depicted are the grand average classification accuracy for CCA and UMM across varying single-trial durations. Here, a bandpass of 6-50\,Hz is used. For a definition of method names, see~\ref{fig:decodingcurve}. The dashed gray line denotes the theoretical chance level (5\%).}
\end{figure*}

In this study, we assessed the performance of two calibration-free methods when applied to \mbox{c-VEP} data. These methods, CCA and UMM, were evaluated as an instantaneous version that classified each new trial without prior information, treating it as the first, and a cumulative version that learned from the insights gained through previously classified trials.

Acknowledging the potential for each method to exhibit distinct responses to varying bandpass filter hyper-parameters, our initial focus involved determining the optimal hyper-parameters for each method. The classification accuracy of 31.5-second trials across different cutoff frequencies and methods is illustrated in Fig.~\ref{fig:bandpass}.

In the highpass analysis (Fig.~\ref{fig:bandpass}, left side), it was evident that CCA is more sensitive to a low highpass value than UMM. For both CCA's instantaneous and cumulative versions, a decline in accuracy was observed for highpass values below 2\,Hz. In contrast, UMM appeared to be less affected. In the pursuit of the highest accuracy, CCA\_ec achieved a classification accuracy of 0.97 at 2\,Hz, CCA\_e1 0.96 at 6\,Hz, UMM\_tcw 0.94 at 4\,Hz, and UMM\_t11 0.89 at 2\,Hz. When adopting a common highpass at 6\,Hz for all methods, the methods still attained these peak performances.

In the lowpass analysis (Fig.~\ref{fig:bandpass}, right side), an inverse trend emerged, revealing that UMM is more sensitive to the lowpass value than CCA. Specifically, UMM achieved a peak performance only when the lowpass value was set no lower than 40\,Hz, whereas for CCA, this peak was already attained at 20\,Hz. In the pursuit of optimal performance, we identified a peak classification accuracy of 0.97 at 20\,Hz for CCA\_ec, 0.96 at 30\,Hz for CCA\_e1, 0.94 at 50\,Hz for UMM\_tcw, and 0.89 at 50\,Hz for UMM\_t11. When applying a common lowpass at 50\,Hz for all methods, they continued to operate at these peak performance levels.

The above mentioned results led to the selection of a common passband set to 6 to 50\,Hz. We then continued analyzing the behavior of the different methods by estimating so-called decoding curves that show the classification accuracy across different amounts of data available in each single-trial, see Fig.~\ref{fig:decodingcurve}.

From these decoding curves, two general trends could be identified. Firstly, the cumulative versions of both methods (CCA\_ec and UMM\_tcw) outperformed their instantaneous counterparts (CCA\_e1 and UMM\_t11). Secondly, overall, the CCA methods (CCA\_ec and CCA\_e1) outperformed the UMM methods (UMM\_tcw and UMM\_t11). For an overview of the classification accuracy at some time points, please see Tab.~\ref{tab:accuracy}.

\begin{table}[h]
    \centering
    \caption{\label{tab:accuracy}\textbf{Classification accuracy.} Listed are the grand average accuracy reached by the four methods at distinct trial durations. }
    \begin{tabular}{lccccc}
        \toprule
        & \textbf{1.05\,s} & \textbf{2.1\,s} & \textbf{4.2\,s} & \textbf{10.5\,s} & \textbf{31.50\,s} \\
        \midrule
        \textbf{CCA\_ec}     & 0.24  & 0.52 & 0.86  & 0.96  & 0.97 \\
        \textbf{CCA\_e1}     & 0.06  & 0.29 & 0.59  & 0.85  & 0.96 \\
        \textbf{UMM\_tcw}    & 0.13  & 0.39 & 0.75  & 0.94  & 0.94 \\
        \textbf{UMM\_t11}    & 0.09  & 0.19 & 0.37  & 0.69  & 0.89 \\
        \bottomrule
    \end{tabular}
\end{table}

The cumulative CCA method (CCA\_ec) achieved an accuracy of 0.24 at the smallest trial duration of 1.05\,s (half a code cycle) and 0.52 at 2.1\,s (one code cycle). At these two early time points, CCA\_ec did not significantly outperform UMM\_tcw, which reached 0.13 and 0.39, respectively. At all further time points, CCA\_ec did significantly surpass UMM\_tcw. At the maximum trial length of 31.5\,s, CCA\_ec achieved a classification accuracy of 0.97, significantly outperforming UMM\_tcw, which achieved 0.94.

The instantaneous CCA method (CCA\_e1) achieved a performance of 0.06 at 1.05\,s, while the instantaneous UMM (UMM\_t11) reached 0.09. At this time point, UMM\_t11 significantly outperformed CCA\_e1. Instead, at any further time point CCA\_e1 significantly surpassed the accuracy of UMM\_t11. At the maximum trial length of 31.5\,s, CCA\_e1 reached a classification of 0.96 which was significantly higher than UMM\_t11 with 0.89. 

The cumulative versions always outperformed the instantaneous version for both CCA and UMM. Notably, the instantaneous CCA (CCA\_e1) performed almost on par with its cumulative version (CCA\_ec) for trials longer than 25.2\,s. Specifically, at 27.3\,s, CCA\_e1 reached an accuracy of 0.95 while CCA\_ec reached 0.97, which was not significantly higher. Also at the 29.4 and 31.5\,s trials, CCA\_ec did not significantly surpass CCA\_e1.


\mysection{discussion}

We introduced UMM to \mbox{c-VEP} BCI, a calibration-free method originating from the ERP domain. We conducted a systematic offline comparison with the conventional c-VEP method employing CCA. Both methods underwent evaluation in an instantaneous manner, classifying each single trial without prior knowledge, as well as in a cumulative way, utilizing previously classified trials as training data. The ultimate goal was to establish a \mbox{c-VEP} BCI with enhanced usability and broader adoption potential by eliminating the initial calibration session.
We gave both methods access to EEG data only that would be available in an online experiment such that we don't expect a major problem for online use. For CCA approaches, online performance was reported in~\cite{thielen2021} where the non-instantaneous version matched a supervised model after a warm-up period. 

Both the conventional method, CCA, and the novel approach, UMM, hinge on the principle that selecting a single symbol from a set of candidate symbols is considerably simpler than the reconstructing the entire stimulus sequence. Using 20 symbols in this study, one has to evalu\-ate 20 candidate stimulus sequences only for a symbol selection. In contrast, the exhaustive reconstruction of the stimulus would invole an exponential growth, reaching  $2^{60}$ potential sequences for a 1-second stimulus at a presentation rate of 60\,Hz.

Without involving a calibration session, the cumulative CCA already reached higher than 90\,\% performance for trials of 5.25\,s, with the cumulative UMM just behind reaching a similar performance at 7.35-second trials. Despite CCA regularly outperforming UMM, astonishingly, these results demonstrate that UMM performs rather well, despite not being optimized for \mbox{c-VEP} data. 

Furthermore, we demonstrated the capability of instantaneous classification, where CCA reached higher than 90\,\% accuracy at 14.70-second trials and UMM 89\,\% at 29.40\,s. Instantaneous decoding does not learn from, or consider in any way, previous trials. Under stationary conditions, this scarcity of data may not allow the model to reach peak performance. On the other hand, non-stationary feature distributions like latency- or amplitude changes of ERP components over time, will likely not affect instantaneous decoding, while models mainly trained on data collected \textit{before} the appearance of such feature drifts may suffer from performance degradation.

Both CCA and UMM make different assumptions and exploit them. These assumptions may be met by different datasets to various degrees. 
For instance, CCA strongly leverages the sequential structure and large overlap between responses to adjacent stimuli in \mbox{c-VEP} datasets. 
Additionally, the event definition used focusses on the target ERP while discerning between ERPs associated with short and long flashes, but more and different event types can easily be implemented. 
Furthermore, UMM maintains its operations within the original EEG feature space, whereas CCA operates in the component space.

In contrast, UMM searches for the stimulus sequence with the largest target to non-target ERP distance, potentially rendering it less susceptible to slow drifts in the data, as evidenced by the highpass analysis. 
Besides, CCA uses the empirical covariance matrix, which can be challenging to estimate with limited data, while UMM employs domain-specific regularization techniques such as shrinkage and a block-Toeplitz covariance matrix~\cite{sosulski2022}.
Lastly, in the non-instantaneous UMM formulation, ERP mean estimates are improved using previously classified trials, by carefully weighting mean updates based on the confidence of each previous trial.
Gaining a comprehensive understanding of the strengths and limitations of both methods bears the potential to develop refined versions tailored to specific characteristics of novel datasets through thoughtful hyper-parameterization.

Essentially, CCA and UMM necessitate a specific stimulus protocol involving repetitions, and both require knowledge about the precise timing and sequence of stimuli within a single trial (i.e., the selection of one symbol). While such information is typically available in BCI protocols using evoked responses, it may not seamlessly extend to other protocols like those based on sensorimotor rhythms. Moreover, these decoding methods are only applicable for benchmarks if the sequence information is provided, as demonstrated in MOABB~\cite{aristimunha2023}. This characteristic classifies the studied CCA and UMM methods as semi-supervised, given that stimulus information is requi\-site, while label information is not required.




\mysection{conclusion}

We showed that both CCA and UMM offer the potential to eliminate the necessity for a calibration session, thereby enhancing the usability for BCI applications, especially when integrated with the \mbox{c-VEP} protocol. These findings mark an initial stride toward combining the robust capabilities of machine learning methods across diverse domains. They inspire the exploration of their cross-application and cross-pollination, unlocking new possibilities for advancing BCI technologies.


\mysection{Acknowledgements}

This work was part of the project `Dutch Brain Interface Initiative' (DBI2) with project number 024.005.022 of the research programme `Gravitation' which is (partly) financed by the Dutch Research Council (NWO).


\mysection{references}
\printbibliography[heading=none]

\end{document}